\documentclass[conference]{IEEEtran}
\usepackage{amsmath,amsfonts}
\usepackage{algorithmic}
\usepackage{array}
\usepackage[caption=false,font=normalsize,labelfont=sf,textfont=sf]{subfig}
\usepackage[group-separator={,},group-minimum-digits=3]{siunitx}
\usepackage{textcomp}
\usepackage{stfloats}
\usepackage{url}
\usepackage{verbatim}
\usepackage{graphicx}
\usepackage{balance}
\usepackage{mathrsfs}
\def\BibTeX{{\rm B\kern-.05em{\sc i\kern-.025em b}\kern-.08em
    T\kern-.1667em\lower.7ex\hbox{E}\kern-.125emX}}
\begin{document}
\title{ \fontsize{21pt}{\baselineskip}\selectfont {An Anti-Jamming Strategy for Disco Intelligent Reflecting Surfaces Based Fully-Passive Jamming Attacks}
}
\author{\IEEEauthorblockN{Huan~Huang\IEEEauthorrefmark{1}, Hongliang~Zhang\IEEEauthorrefmark{2}, Yi~Cai\IEEEauthorrefmark{1}, A.~Lee~Swindlehurst\IEEEauthorrefmark{3}, and~Zhu~Han\IEEEauthorrefmark{4}}
\IEEEauthorblockA{\IEEEauthorrefmark{1} Electronic Information School, Soochow University, Suzhou, China}
\IEEEauthorblockA{\IEEEauthorrefmark{2} School of Electronics, Peking University, Beijing, China}
\IEEEauthorblockA{\IEEEauthorrefmark{3} Department of Electrical and Computer Engineering, University of Houston, Houston, USA}
\IEEEauthorblockA{\IEEEauthorrefmark{4} Center for Pervasive Communications and Computing, University of California, Irvine, USA}
\IEEEauthorblockA{Email: hhuang1799@gmail.com, hongliang.zhang92@gmail.com, yicai@ieee.org, swindle@uci.edu, hanzhu22@gmail.com}
}
\maketitle
\begin{abstract}
Emerging intelligent reflecting surfaces (IRSs) significantly improve system performance, while also pose a huge risk for physical layer security. A disco IRS (DIRS), i.e., an illegitimate IRS with random time-varying reflection properties, can be employed by an attacker to actively age the channels of legitimate users (LUs). Such active channel aging (ACA) generated by the DIRS-based fully-passive jammer (FPJ) can be applied to jam multi-user multiple-input single-output (MU-MISO) systems without relying on either jamming power or LU channel state information (CSI). To address the significant threats posed by the DIRS-based FPJ, an anti-jamming strategy is proposed that requires only the statistical characteristics of DIRS-jammed channels instead of their CSI. Statistical characteristics of DIRS-jammed channels are first derived, and then the anti-jamming precoder is given based on the derived statistical characteristics. Numerical results are also presented to evaluate the effectiveness of the proposed anti-jamming precoder against the DIRS-based FPJ.
\end{abstract}

\begin{IEEEkeywords}
Physical layer security, intelligent reflecting surface, transmit precoding, jamming suppression.
\end{IEEEkeywords}

\section{Introduction}\label{Intro}
Due to the open communication environment in wireless broadcast and superposition channels, wireless networks are vulnerable to malicious attacks such as jamming (also known as DoS-type attacks) and eavesdropping~\cite{PLSsur1,DoSsur1}. 
Jamming attacks can be launched by an active jammer (AJ), which inflicts intentional jamming/interference to block a communication between the wireless access point (AP) and its legitimate users (LUs). Generally, physical-layer AJs can be divided into the following categories~\cite{DoSsur1}: constant AJs, intermittent AJs, reactive AJs, and adaptive AJs.

Recently, intelligent reflecting surfaces (IRSs), a promising wireless technology for future 6G communications, have been proposed to reflect electromagnetic waves in a controlled manner~\cite{IRSsur1,IRSsur2}. 
Previous works have focused mainly on the introduction of legitimate IRSs to improve certain performance metrics such as energy efficiency (EE)~\cite{MyEE}, spectrum efficiency (SE)~\cite{MySE}, or cell coverage~\cite{PostDrRenQ}. 
However, some works have pointed out that illegitimate IRSs can impose a significant impact on wireless networks because the illegitimate IRSs are difficult to detect due to their passive nature~\cite{IIRSSur}.
Fortunately, active jamming of the LUs by the illegitimate IRS requires channel state information (CSI) of all LU channels involved.
If the illegitimate IRS wants to acquire LU CSI, it needs to estimate the CSI jointly with the legitimate AP and LUs, and thus it is very difficult to implement in practice.

An interesting fully-passive jammer (FPJ)~\cite{DIRSVT,DIRSTWC} has been proposed to launch jamming attacks on LUs with neither LU CSI nor jamming power, where an illegitimate IRS with random phase shifts, referred to as a ``disco" IRS (DIRS), is used to actively age the LUs' channels to cause serious active channel aging (ACA) interference.
Classical anti-jamming approaches~\cite{AntiJammingSurv}, such as spread spectrum and frequency-hopping techniques, can not be used against this type of FPJ.
This is because the source of jamming attacks launched by the FPJ comes from the legitimate AP transmit signals, and therefore always has the same characteristics (such as the carrier frequency) as the transmit signals.

In addition, the jamming attacks from the DIRS-based FPJ, i.e., the ACA interference, can not be mitigated using multi-input multi-output (MIMO) interference cancellation~\cite{PTDTSyn}. MIMO interference cancellation is effective for DIRS-based ACA interference only if the CSI of both the LU and DIRS-jammed channels is known by the legitimate AP~\cite{AntiJammingSurv,PTDTSyn}. However, the DIRS electromagnetic responses (such as phase shifts and amplitudes) are randomly generated~\cite{DIRSVT,DIRSTWC}. In summary, there is no effective anti-jamming approach available to counteract the destructive DIRS-based ACA interference imposed by the FPJ.

In this paper, we investigate DIRS-based ACA interference caused by an FPJ and propose an effective anti-jamming strategy.
The main contributions are summarized as follows:
\begin{itemize}
\item We consider a practical IRS model in which the phase shifts of the DIRS reflecting elements are discrete and the amplitudes are a function of their corresponding phase shifts. Based on this IRS model, we describe the DIRS-based FPJ, which initiates jamming attacks through the DIRS-based ACA interference and requires no additional jamming power or knowledge of the LU CSI. 
\item To address the significant threats posed by the DIRS-based FPJ, we develop an anti-jamming strategy that requires only the statistical characteristic of DIRS-jammed channels and avoids requiring their instantaneous CSI, which is impractical to obtain. We first derive a statistical characteristic of the DIRS-jammed channels, and then prove that the developed anti-jamming precoder can achieve the maximum signal-to-jamming-plus-noise ratio (SJNR).
\end{itemize}

The rest is organized as follows. In Section~\ref{FPJ}, we consider a practical IRS model and describe the general mode of the DIRS-based FPJ. Then, we define the SJNR optimization metric and present the channel model.
In Section~\ref{StatisticalCha}, we first derive a statistical characteristic of the DIRS-jammed channels. Then, we develop an anti-jamming precoder based on the derived statistical characteristic and prove that this precoder can achieve the maximum SJNR.
Simulation results are provided in Section~\ref{ResDis} to show the effectiveness of the proposed anti-jamming precoder against the DIRS-based FPJ. Finally, the main conclusions are given in Section~\ref{Conclu}.

We employ bold capital type for a matrix, e.g., $\boldsymbol{\Phi}\!_{{D\!T}}$, small bold type for a vector, e.g., $\boldsymbol{w}\!_{R\!P\!T,k}$, and italic type for a scalar, e.g., $K$. The superscripts $(\cdot)^{H}$ and $(\cdot)^{-1}$ represent the Hermitian transpose and the inverse. The symbols $|\cdot|$ and $\|\cdot\|$ denote the absolute value and the Frobenius norm.
\section{System Statement}\label{FPJ}
\subsection{Disco-IRS-Based Fully-Passive Jammer}\label{DIRSFPJ}
\begin{figure}[!t]
\centering
\includegraphics[scale=0.68]{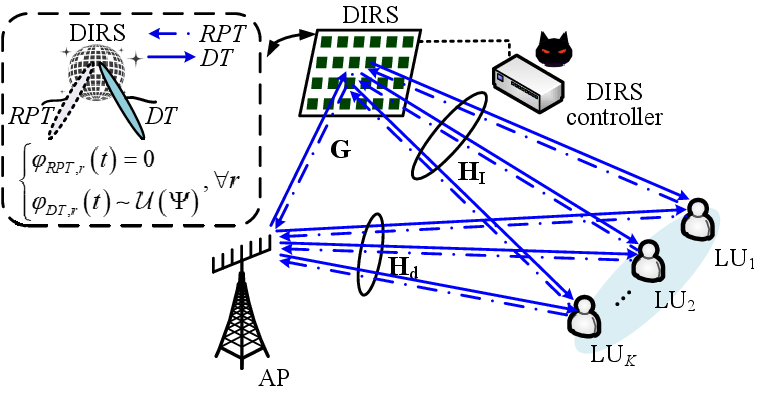}
\caption{Illustration of fully passive jamming (FPJ) implemented by disco intelligent reflecting surface (DIRS) based active channel aging in an MU-MISO system. RPT: reverse pilot transmission; DT: data transmission.}
\label{fig1}
\end{figure}

Fig.~\ref{fig1} diagrammatically demonstrates an MU-MISO system jammed by the DIRS-based FPJ. Specifically, the legitimate AP has $N_{\rm A}$ antennas and communicates with $K$ single-antenna LUs denoted by ${\rm {LU}}_{1}, {\rm {LU}}_{2}, \cdots, {\rm {LU}}_{K}$. A DIRS consisting of $N_{\rm D}$ reflecting elements with one-bit quantized phase is placed relatively close to the AP$\footnote{Many existing performance-enhancing IRS-based systems assume that legitimate IRSs are placed close to the users to maximize system performance~\cite{MyEE,MySE,PostDrRenQ}. However, here we make the more robust assumption that the DIRS controller has no information about the LUs~\cite{DIRSVT,DIRSTWC}. Therefore, the DIRS is placed relatively close to the AP to maximize the impact of the DIRS.}$ to jam the LUs.

In an MU-MISO system, the channel coherence time consists of two phases, i.e., a \emph{reverse pilot transmission (RPT)} phase and a \emph{data transmission (DT)} phase. In general, the wireless channels in a traditional MU-MISO system are assumed to remain unchanged during the channel coherence time. Therefore, the CSI is estimated during the \emph{RPT} phase, and the downlink precoding is then designed based on the obtained CSI and used to transmit signals to the LUs during the \emph{DT} phase. However, the work in~\cite{DIRSVT,DIRSTWC} has shown that the DIRS can be used to actively age the channels, i.e., create rapidly changing wireless channels within the channel coherence time. The DIRS-based ACA differs from channel aging (CA) in traditional MU-MISO systems~\cite{ChanAge}. In fact, CA is CSI inaccuracy due to time variation of channels and computation delays, and it can not be actively introduced and controlled.
\subsubsection{Data Transmission} During the \emph{DT} phase, the signal received at ${\rm{LU}}_k$ is given by
\begin{alignat}{1}
\nonumber
{y_k} &= \boldsymbol{h}\!_{D\!T,k}^H\!\sum_{u = 1}^K {{\boldsymbol{w}\!_{R\!P\!T,u}}{s_u}}  + {n_k}\\
&= \!\left(\! { {\boldsymbol{h}_{{\rm{I}},k}^H{{\bf{\Phi}}\!_{D\!T}}{{\bf{G}}}} + \boldsymbol{h}_{{\rm{d}},k}^H}\! \right)\!\!\sum_{u = 1}^K {{\boldsymbol{w}\!_{R\!P\!T,u}}{s_u}}  + {n_k},
\label{RecSig}
\end{alignat}
where the transmitted signal $s_u\in \mathbb{C}$ satisfies ${\mathbb{E}}\!\left[ {{s_u}s_u^H} \right] \!= \!1$. In addition, $\boldsymbol{h}\!^H_{D\!T,k} \!\in\! {\mathbb{C}^{1\!\times\!{N_{\rm{A}}}}}$ represents the combined channel between the legitimate AP and ${\rm{LU}}_k$, $\boldsymbol{h}_{{\rm{I}},k} \in\mathbb{C}^{N_{\rm{D}}\!\times\! 1}$ represents the channel between the DIRS and ${\rm{LU}}_k$, ${\bf{G}}=\left[\boldsymbol{g}_{1},\cdots,\boldsymbol{g}_{{N}\!_{\rm A}}\right] \in\mathbb{C}^{N_{\rm{D}}\!\times\! N_{\rm{A}}}$ represents the channel between the legitimate AP and the DIRS, and $\boldsymbol{h}_{{\rm{d}},k} \in\mathbb{C}^{N_{\rm{A}}\!\times\! 1}$ represents the direct channel between the legitimate AP and ${\rm{LU}}_k$. Furthermore, $n_k$ denotes additive white Gaussian noise (AWGN) with zero mean and variance $\sigma^2$, i.e., $n_k\sim \mathcal{CN}\left(0,\sigma^2\right)$. For ease of presentation, the overall DIRS-LU channel ${\bf{H}}_{\rm I}$ and the overall LU direct channel ${\bf{H}}_{\rm d}$ are denoted as ${\bf{H}}_{\rm I} = \left[\boldsymbol{h}_{{\rm{I}},1},\cdots,\boldsymbol{h}_{{\rm{I}},K}\right]$ and ${\bf{H}}_{\rm d} = \left[\boldsymbol{h}_{{\rm{d}},1},\cdots,\boldsymbol{h}_{{\rm{d}},K}\right]$, respectively. Furthermore, the DIRS-jammed channel ${\bf{H}}_{\rm D}^{DT}$ is denoted as ${\bf{H}}_{\rm D}^{DT} = {{\bf{H}}_{{\rm{I}}}^H{{\bf{\Phi}}\!_{D\!T}}{{\bf{G}}}} = \left[\boldsymbol{h}\!^{D\!T}_{{\rm{D}},1},\cdots,\boldsymbol{h}\!^{D\!T}_{{\rm{D}},K}\right]$.

In~\eqref{RecSig}, $\boldsymbol{\Phi}\!_{D\!T}\!=\!{\rm{diag}}\!\left({\boldsymbol{\varphi}}\!_{D\!T}\!(t)\right)\in\mathbb{C}^{N_{\rm{D}}\!\times\! N_{\rm{D}}}$ denotes the passive beamformer of the DIRS during the $DT$ phase.
The reflecting vector ${\boldsymbol{\varphi}}\!_{D\!T} (t)$ is randomly generated and given by ${\boldsymbol{\varphi}}\!_{D\!T}(t) = \! \left[ {{\beta\!_{D\!T,1}}(t){e^{j{\varphi\!_{D\!T,1}}(t)}}, \cdots ,{\beta\!_{D\!T,{N\!_{\rm{D}}}}\!(t)}{e^{j{\varphi\!_{{N\!_{\rm{D}}}}}\!(t)}}} \right]$,
where ${\beta\!_{D\!T,r}}(t)$ and ${\varphi\!_{D\!T,r}}(t)$ $(r=1,\cdots, N\!_{\rm D})$ denote the amplitude and phase shift of the $r$-th DIRS reflecting element, respectively.
An IRS is an ultra-thin surface inlaid with multiple sub-wavelength reflecting elements whose electromagnetic responses (such as phase shifts and amplitudes) are controlled by simple programmable PIN diodes~\cite{IRSsur2}.
Based on the ON/OFF behavior of PIN diodes, only a few discrete phase shifts are generated by an IRS.

For a DIRS with $b$-bit quantized phase shifts, let $\Psi  = \left\{ {{\theta _1}, \cdots ,{\theta _{{2^b}}}} \right\}$ represent the set of discrete phase shift values. Assume that the DIRS phase shifts are randomly selected from $\Psi$, i.e., ${\varphi\!_{ D\!T,r}}(t)\sim {\cal R}\!\left(\Psi\right)$. The gain values ${\beta\!_{D\!T,r}}(t)$ are a function of ${\varphi\!_{D\!T,r}}(t)$~\cite{IRSsur1}, which we express as ${\beta\!_{D\!T,r}}(t) = {\cal F}\left( {{\varphi\!_{D\!T,r}}(t)} \right)$. We denote the set of all possible amplitude values as $\Omega  = {\cal F}\left( \Psi \right) = \left\{ {{\kappa _1}, \cdots ,{\kappa _{{2^b}}}} \right\}$. Note that, in practice, the independent DIRS controller does not have sufficient computing power to generate complex distributions for the DIRS reflecting vector ${\boldsymbol{\varphi}}\!_{D\!T} (t)$~\cite{DIRSVT,DIRSTWC}. 
Therefore, we assume that the illegitimate DIRS jams the MU-MISO system with reflecting phase shifts that follow a uniform distribution, i.e., ${{\varphi\!_{D\!T,r}}(t)} \sim {\cal U}\left(\Psi\right)$.
\subsubsection{Reverse Pilot Transmission}
To optimize the multi-user active beamforming ${\bf{W}}\!_{R\!P\!T} \!=\! \left[{\boldsymbol{w}\!_{R\!P\!T,1}},\cdots,{\boldsymbol{w}\!_{R\!P\!T,K}}  \right]\in \mathbb{C}^{N\!_{\rm A}\times K}$, the legitimate AP needs to obtain the LUs' CSI using pilot-based channel estimation after the $RPT$ phase. Specifically, the LUs send pilot symbols to the legitimate AP, and the AP then estimates the channel using, for example, a traditional solution such as the least squares (LS) algorithm.

During the $RPT$ phase, the DIRS reflecting vector is denoted by ${\boldsymbol{\varphi}}\!_{R\!P\!T} (t)$. Similar to~\cite{DIRSTWC}, the DIRS remains silent during the $RPT$ phase, i.e.,  ${\boldsymbol{\varphi}}\!_{R\!P\!T} (t) = {\boldsymbol{0}}$, which means that the wireless signals are perfectly absorbed by the illegitimate DIRS. 
Consequently, the overall multiuser channel estimated by the legitimate AP is written as
\begin{equation}
{{\bf{H}}\!_{{R\!P\!T}}^H}  = \left[{{\boldsymbol{h}}\!_{{R\!P\!T},1}}, \cdots, {{\boldsymbol{h}}\!_{{R\!P\!T},K}}\right]^H = {\bf{H}}_{\rm d}^H .
\label{HTotalRPT}
\end{equation}

Generally, the aim of the multi-user active beamforming optimization is to maximize the signal power and minimize the interference leakage. Based on the CSI obtained for ${{\bf{H}}\!_{{R\!P\!T}}}$, a widely-used approach is the zero-forcing beamforming (ZFBF)~\cite{ZFBF}, which results in zero interference leakage. Specifically, ${\bf{W}}\!_{R\!P\!T}$ calculated with the ZFBF algorithm is given by
\begin{equation}
{{\bf{W}}\!_{{R\!P\!T}}} = {{{{\bf{H}}\!_{{R\!P\!T}}}{{\left( {{\bf{H}}\!_{{R\!P\!T}}^H{{\bf{H}}\!_{{R\!P\!T}}}} \right)}^{ - 1}}{{\bf{P}}^{ \frac{1}{2}}}}} =\left[{{\boldsymbol{w}}\!_{{R\!P\!T},1}}, \cdots, {{\boldsymbol{w}}\!_{{R\!P\!T},K}}\right],
\label{ZF}
\end{equation}
where $\|{{\boldsymbol{w}}\!_{{R\!P\!T},k}}\| = \sqrt{p_{k}} $ and $p_k$ represents the transmit power of ${\rm{LU}}_k$. The total transmit power $P_0$ at the legitimate AP satisfies the constraint that $\sum\nolimits_{k = 1}^K {{p_k}}  \le {P_0}$. For simplicity, we assume that $p_{k} = \frac{P_{0}}{K}, \forall k$.
\subsubsection{Active channel Aging}
In a traditional MU-MISO system, a wireless channel is assumed to be essentially invariant during its coherence time. Each coherence time includes an $RPT$ phase and a much longer $DT$ phase, and thus the CSI obtained in the $RPT$ phase is assumed to be the same as in the $DT$ phase.

However, an IRS offers the ability to actively age the channel~\cite{DIRSVT,DIRSTWC} and produces a situation where the CSI obtained in the $RPT$ phase is different from that in the $DT$ phase. As a result, serious DIRS-based ACA interference is introduced. The SJNR for LU$_k$ quantifies the DIRS-based ACA interference, and based on~\eqref{RecSig} is given by~\cite{RefSLNR,RefSLNRadd}:
\begin{equation}
{\eta _k} = \frac{{\mathbb{E}}\!\!\left[{{{\left| {\boldsymbol{h}\!_{D\!T,k}^H{\boldsymbol{w}\!_{R\!P\!T,k}}} \right|}^2}}\right]}{{\sum\limits_{u \ne k} \!{\mathbb{E}}\!\!\left[{{{\left| {\boldsymbol{h}\!_{D\!T,u}^H{\boldsymbol{w}\!_{R\!P\!T,k}}} \right|}^2} }\right] }+ {\sigma^2} }.
\label{eqSLNR}
\end{equation}

\subsection{Channel Model}\label{ChaMod}
Since the DIRS is deployed relatively close to the legitimate AP, the AP-DIRS channel $\bf{G}$ is assumed to follow Rician fading~\cite{AORIS}. Meanwhile, the overall DIRS-LU channel ${\bf{H}}_{\rm I}$ and the overall LU direct channel ${\bf{H}}_{\rm d}$ are assumed to follow Rayleigh fading~\cite{AORIS}. Mathematically, $\bf{G}$ is modelled as
\begin{equation}
{{\bf{G}}}  =  {\sqrt{{\mathscr{L}}_{\rm G}}}  \left(\!{\sqrt {\frac{{\varepsilon}}{{1\!+\!\varepsilon}}} {\bf{G}}^{{\rm{LOS}}}\!+\!\sqrt {\frac{1}{{1 + \varepsilon}}} {\bf{G}}^{{\rm{NLOS}}}}\!\!\right),
\label{ChannelG}
\end{equation}
where ${\mathscr{L}}_{\rm G}$ represents the large-scale channel fading for $\bf{G}$, $\varepsilon$ is the Rician factor, and ${\bf{G}}^{{\rm{LOS}}}$ and ${\bf{G}}^{{\rm{NLOS}}}$ denote the line-of-sight (LOS) and non-line-of-sight (NLOS) component of ${{\bf{G}}}$. More specifically, ${\bf{G}}^{{\rm{NLOS}}}$ has i.i.d. elements given by $\left[{\bf{G}}^{{\rm{NLOS}}}\right]_{r,n} \sim \mathcal{CN}\left(0,1\right)$. Moreover, the DIRS is deployed close to the legitimate AP, and thus the element $ \left[{\bf{G}}^{{\rm{LOS}}} \right]_{r,n}$ is characterized by the near-field model~\cite{NearfieldMo} as follows:
\begin{equation}
\left[{\bf{G}}^{{\rm{LOS}}}\right]_{r,n} = {e^{ - j\frac{{2\pi }}{\lambda }\left( {{D_n^r} - {D_n}} \right)}},
\label{GLOS}
\end{equation}
where $\lambda$ represents the wavelength of the transmit signals, and ${D_n^r}$ and ${D_n}$ denote the distance between the $n$-th antenna and the $r$-th DIRS reflecting element, and the distance between the $n$-th antenna and the centre (origin) of the DIRS, respectively. We assume that both the distance between adjacent antennas and the distance between adjacent DIRS reflecting elements are $d = {\lambda}/2$.

The channels ${\bf{H}}_{\rm I}$ and ${\bf{H}}_{\rm d}$ are mathematically modelled as
\begin{alignat}{1}
&{{\bf{H}}_{\rm{I}}} = {\widehat {\bf{H}}_{\rm{I}}}{{\bf{D}}_{\rm I}^{{1 \mathord{\left/
 {\vphantom {1 2}} \right.
 \kern-\nulldelimiterspace} 2}}} = \left[ {{\sqrt{{{\mathscr{L}}_{{\rm I},1}}}}{{\widehat {\boldsymbol{h}}}_{{\rm I},1}}, \cdots ,{\sqrt{{{\mathscr{L}}_{{\rm I},K}}}}{{\widehat {\boldsymbol{h}}}_{{\rm I},K}}} \right], \label{HIkeq}\\
&{{\bf{H}}_{\rm{d}}} = {\widehat {\bf{H}}_{\rm{d}}}{{\bf{D}}_{\rm d}^{{1 \mathord{\left/
 {\vphantom {1 2}} \right.
 \kern-\nulldelimiterspace} 2}}} = \left[ {{\sqrt{{{\mathscr{L}}_{{\rm d},1}}}}{{\widehat {\boldsymbol{h}}}_{{\rm d},1}}, \cdots ,{\sqrt{{{\mathscr{L}}_{{\rm d},K}}}}{{\widehat {\boldsymbol{h}}}_{{\rm d},K}}} \right],
\label{Hdkeq}
\end{alignat}
where the elements of the $K\times K$ diagonal matrices ${\bf{D}}_{\rm I} = {\rm{diag}}\left({{\mathscr{L}}_{{\rm I},1}}, \cdots,{{\mathscr{L}}_{{\rm I},K}}\right)$ and ${\bf{D}}_{\rm d} = {\rm{diag}}\left({{\mathscr{L}}_{{\rm d},1}},\cdots,{{\mathscr{L}}_{{\rm d},K}}\right)$ represent the large-scale channel fading. Moreover, the i.d.d. elements of ${\widehat {\bf{H}}_{\rm{I}}}$ and ${\widehat {\bf{H}}_{\rm{d}}}$ are defined as $[{\widehat {\bf{H}}_{\rm{I}}}]_{r,k}, [{\widehat {\bf{H}}_{\rm{d}}}]_{n,k} \sim \mathcal{CN}\left(0,1\right)$, $r = 1, \cdots, N_{\rm D}$, $n=1, \cdots, N_{\rm A}$, and $k = 1,\cdots,K$.
\setlength{\topmargin}{-0.681 in}
\section{An Anti-Jamming Strategy for Disco-IRS-Based Fully-Passive Jammers}\label{StatisticalCha}
As mentioned in Section~\ref{Intro}, it is unrealistic for the legitimate AP to have the
knowledge of the DIRS-jammed channel ${\bf{H}}_{\rm D}^{D\!T}$.
In order to develop an anti-jamming precoder for the DIRS-based FPJ presented in Section~\ref{FPJ}, we derive the following statistical characteristic of ${\bf{H}}_{\rm D}^{D\!T}$, i.e., Proposition~\ref{Proposition1}.
\newtheorem{proposition}{Proposition}
\begin{proposition}
\label{Proposition1}
The i.d.d. elements of ${\bf{H}}_{\rm D}^{DT}$ converge in distribution to $\mathcal{CN}\!\left( {0,  {{{\mathscr{L}}\!_{{\rm G}}}{{\mathscr{L}}\!_{{\rm I},k}}{N\!_{\rm D}}\delta^2 } } \right)$ as $N_{\rm D} \to \infty$, i.e.,
\begin{equation}
{\left[ {{\bf H}_{\rm{D}}^{DT}} \right]_{k,n}} \mathop  \to \limits^{\rm{d}} \mathcal{CN}\!\left( {0,  {{{\mathscr{L}}\!_{{\rm G}}}{{\mathscr{L}}\!_{{\rm I},k}}{N\!_{\rm D}}\delta^2 } } \right), \forall k,n,
\label{HDSta}
\end{equation}
where $\delta^2 = \frac{{\sum\nolimits_{i = 1}^{{2^b}} {\kappa _i^2} }}{{{2^b}}}$.
\end{proposition}

\begin{IEEEproof}
 See Appendix~\ref{AppendixA}.
\end{IEEEproof}
It is worth noting that the DIRS-based FPJ is deployed in practice with a large number of reflecting elements, i.e., $N\!_{\rm D} \gg 1$.
According to the previous work~\cite{DIRSTWC}, the DIRS must be equipped with a massive number of elements in order to launch significant jamming attacks since the large-scale channel fading in the DIRS-jammed channel ${\bf{H}}_{\rm D}^{DT}$ is much more severe than that in the overall LU direct channel ${\bf{H}}_{\rm d}$.

According to Proposition~\ref{Proposition1}, the legitimate AP can use the anti-jamming precoder given in Theorem~\ref{Theorem1} to maximize the SJNR expressed by~\eqref{eqSLNR}.
\newtheorem{theorem}{Theorem}
\begin{theorem}
\label{Theorem1}
The optimal precoder for LU$_k$ to mitigate the jamming attacks launched by the DIRS-based FPJ, i.e., to maximize the SJNR $\eta_k$, is given by
\begin{equation}
{\boldsymbol{w}\!_{{\rm{Anti}},k}} \propto \max.{\rm{eigenvector}}\left({\bf A}_k\right),
\label{AntiJamm}
\end{equation}
where
\begin{equation}
{\bf A}_k = \frac{{{\boldsymbol{h}_{{\rm{d}},k}}\boldsymbol{h}_{{\rm{d}},k}^H + {{{\mathscr{L}}\!_{{\rm G}}}{{\mathscr{L}}\!_{{\rm I},k}}{N\!_{\rm D}}\delta^2 }{{\bf{I}}\!_{N\!_{\rm A}}}}}{{ {{{\widetilde {\bf{H}}}_{{\rm{d}},k}}\widetilde {\bf{H}}_{{\rm{d}},k}^H + \! \left( {\frac{{{\sigma ^2}}}{{{p_k}}} + \!{\sum\limits_{u \ne k}{{\mathscr{L}}\!_{{\rm G}}}{{\mathscr{L}}\!_{{\rm I},u}}{N\!_{\rm D}}\delta^2 }} \right){{\bf{I}}\!_{N\!_{\rm A}}}}  }},
\label{AntiJammMat}
\end{equation}
$\max.{\rm{eigenvector}}\left({\bf A}_k\right)$ denotes the eigenvector of ${\bf A}_k$ associated with the largest eigenvalue, and ${{\widetilde {\bf{H}}}_{{\rm{d}},k}} = \left[{\boldsymbol{h}_{{\rm{d}},1}},\cdots,{\boldsymbol{h}_{{\rm{d}},k-1}},{\boldsymbol{h}_{{\rm{d}},k+1}},\cdots,{\boldsymbol{h}_{{\rm{d}},K}} \right]$.
\end{theorem}

\begin{IEEEproof}
In~\eqref{eqSLNR}, the overall LU direct channel ${\bf{H}}_{\rm d}$ and the multi-user active beamforming ${\bf{W}}\!_{R\!P\!T}$ are fixed during the channel coherence time.
Consequently, the SJNR at ${\rm {LU}}_{k}$ in~\eqref{eqSLNR} reduces to
\begin{equation}
{\eta _k} \!=\! \frac{ {{{\left| {\boldsymbol{h}\!_{{\rm d},k}^H{\boldsymbol{w}\!_{R\!P\!T,k}}} \right|}^2}}\!\! + \! {{{  {\boldsymbol{w}\!_{R\!P\!T,k}^H}{\mathbb{E}}\!\left[\boldsymbol{h}_{{\rm D},k}^{D\!T}({\boldsymbol{h}_{{\rm D},k}^{D\!T}})^H\right]\!{\boldsymbol{w}\!_{R\!P\!T,k}}   } }}  }{\!{\sum\limits_{u \ne k}\! \!\!\left(\! {{{\left| {\boldsymbol{h}\!_{{\rm d},u}^H{\!\boldsymbol{w}\!_{R\!P\!T,k}}} \right|}^2}}\!\! + \!  {{{  {\boldsymbol{w}\!_{R\!P\!T,k}^H}{\mathbb{E}}\!\left[\boldsymbol{h}_{{\rm D},u}^{D\!T}\!(\!{\boldsymbol{h}_{{\rm D},u}^{D\!T}})^H\!\right]\!{\boldsymbol{w}\!_{R\!P\!T,k}}   } }} \! \!\right) }\!\!+\! {\sigma^2} }.
\label{eqSLNRred}
\end{equation}

According to~\eqref{eqSLNRred} and Proposition~\ref{Proposition1}, the following equation can be generated, i.e.,
\begin{equation}
{\eta _k} = \frac{ {{{  {\boldsymbol{w}\!_{R\!P\!T,k}^H} { \widehat {\bf{H}}}\!_{D\!T,k} {{\boldsymbol{w}\!_{R\!P\!T,k}}}  } }} }{ {{{  {\boldsymbol{w}\!_{R\!P\!T,k}^H} { \widehat {\widetilde {\bf H}} }\!_{D\!T,k} {{\boldsymbol{w}\!_{R\!P\!T,k}}}  } }}  },
\label{eqSLNRequMore}
\end{equation}
where ${ \widehat {\bf{H}}} _{DT,k} = {{ { {\boldsymbol{h}_{{\rm{d}},k}}\boldsymbol{h}_{{\rm{d}},k}^H  +  \left( {  { {{\mathscr{L}} _{{\rm G}}}{{\mathscr{L}}_{{\rm I},k}}{N _{\rm D}}\delta^2 }} \right){{\bf{I}} _{N\!_{\rm A}}}}  }}$ and ${ \widehat {\widetilde {\bf H}} }\!_{D\!T,k} \!=\!  {{ {{{\widetilde {\bf{H}}}_{{\rm{d}},k}}\widetilde {\bf{H}}_{{\rm{d}},k}^H \!+\! \left( {\frac{{{\sigma ^2}}}{{{p_k}}} \!+\! {\sum\nolimits_{u \ne k}{{\mathscr{L}}\!_{{\rm G}}}{{\mathscr{L}}\!_{{\rm I},u}}{N\!_{\rm D}}\delta^2 }} \right)\!{{\bf{I}}_{N\!_{\rm A}}}}  }}$.

Using the Rayleigh-Ritz quotient~\cite{RefSLNR} we have
\begin{equation}
{\eta _k} \le {\lambda _{\max }}\!\left(\! { { \widehat {\bf{H}}}\!_{D\!T,k},{ \widehat {\widetilde {\bf H}} }\!_{D\!T,k} } \right).
\label{eqRayRitz}
\end{equation}
Based on~\eqref{eqRayRitz}, the anti-jamming precoder for LU$_k$ that maximizes the SJNR $\eta_k$ is given by
\begin{equation}
{\boldsymbol{w}\!_{{\rm{Anti}},k}} = \sqrt{p_k}\frac{\max.{\rm{eigenvector}}\left({\bf A}_k\right)}{\|\max.{\rm{eigenvector}}\left({\bf A}_k\right)\|}.
\label{AntiJammPrecoding}
\end{equation}
\end{IEEEproof}
As a result, the legitimate AP can exploit the anti-jamming precoder in Theorem~\ref{Theorem1} to mitigate the jamming attacks launched by the DIRS-based FPJ~\cite{DIRSVT,DIRSTWC}. 
\setlength{\topmargin}{-0.721 in}
\section{Simulation Results and Discussion}\label{ResDis}
In this section, we present numerical results to evaluate the effectiveness of the proposed anti-jamming precoder against the DIRS-based FPJ~\cite{DIRSVT,DIRSTWC}. We consider an MU-MISO system where the legitimate AP is equipped with 16 antennas and communicates with eight LUs, i.e., $N_{\rm A} =16$ and $K=8$. Meanwhile, the MU-MISO system is jammed by the DIRS-based FPJ described in Section~\ref{DIRSFPJ}, where the DIRS has $(32\times 32)$ reflecting elements, i.e., $N_{\rm D} = 1024$.
We further consider a one-bit DIRS that is easy to implement in practice, where $\Psi  = \left\{ {\frac{\pi}{9},\frac{6\pi}{5}} \right\}$ and $\Omega  = {\cal F}\left( \Psi \right) = \left\{ {0.8,1 } \right\}$~\cite{IRSsur1}. In other words, the $r$-th element of the reflecting vector ${\boldsymbol{\varphi}}\!_{D\!T} (t)$ satisfies ${\beta\!_{D\!T,r}}(t){e^{j{\varphi\!_{D\!T,r}}(t)}} \in \left\{ 0.8{e^{j\frac{\pi}{9}}},{e^{j\frac{6\pi}{5}}} \right\}$. Consequently, the variance in~\eqref{HDSta} is calculated as $\delta^2 = 0.82$.

The legitimate AP and the DIRS are located at (0m, 0m, 2m) and (2m, 0m, 2m), respectively. The LUs are randomly distributed in a circular region centred at (0m, 180m, 0m) with a radius of 10m. According to the 3GPP propagation model~\cite{3GPP}, the propagation parameters of the wireless channels modelled in~Section~\ref{ChaMod} are described as follows: ${\mathscr{L}}_{\rm G} \!=\!35.6\!+\!22{\log _{10}}({d_{i}})$ and $ {\mathscr{L}}_{k} = {\mathscr{L}}_{{\rm I},k}\!=\!32.6\!+\!36.7{\log _{10}}({d_{{i}}})$, where $d_{i} \in \left\{d_{\rm G}, d_{{\rm I},k}, d_{{\rm d},k}\right\}$ is the propagation distance. Moreover,  the AWGN variance is $\sigma^2\!=\!-170\!+\!10\log _{10}\left(BW\right)$ dBm, where the transmission bandwidth is given by $BW=180$ kHz. 

Herein, we illustrate the average rate per LU$\footnote{The rate per LU here is defined as $\frac{{\sum\nolimits_{k = 1}^K {{{\log }_2}\left( {1 + {\eta _k}} \right)} }}{K}$.}$ achieved by the following benchmarks:  the rate resulting from an MU-MISO system without jamming attacks (W/O Jamming)~\cite{ZFBF}; the rate resulting from an MU-MISO system using the proposed anti-jamming precoder presented in Section~\ref{StatisticalCha} (Proposed Anti-Jamming); the rate resulting from an MU-MISO system under the DIRS-based fully-passive jamming attacks~\cite{DIRSVT,DIRSTWC} (W/FPJ); and the rate resulting from an MU-MISO system under active jamming attacks. In the latter case, two cases are considered with an AJ located at (2m, 0m, 2m) broadcasting jamming signals with -10 dBm jamming power (W/ $P_{\rm J} = -10$ dBm) and -6 dBm jamming power (W/ $P_{\rm J} = -6$ dBm).

\begin{figure}[!t]
\centering
\includegraphics[scale=0.66]{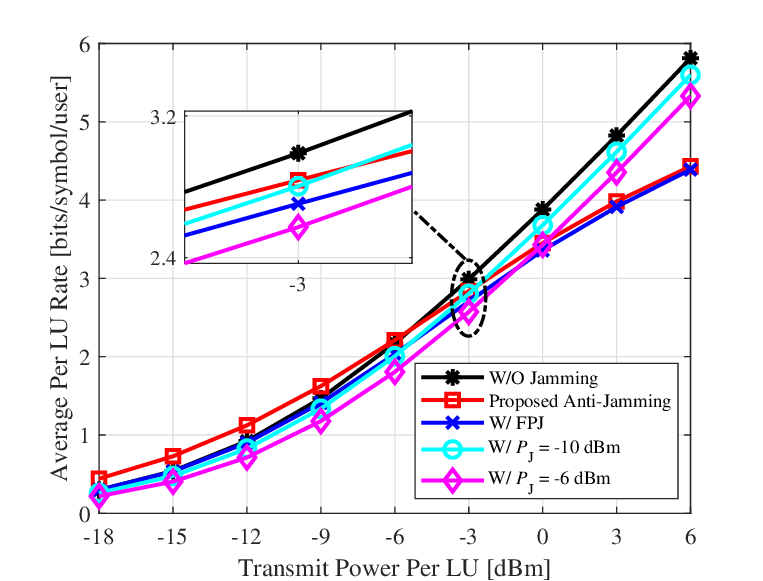}
\caption{Average rate of each legitimate user (LU) vs. transmit power of each LU for different benchmarks.}
\label{Resfig1}
\end{figure}

Fig.~\ref{Resfig1} illustrates the relationship between the average rate per LU and the transmit power per LU. Compared to the average rate per LU obtained W/O Jamming, the results for the proposed anti-jamming precoder are better in the low power domain. This is because the anti-jamming precoder given in Theorem~\ref{Theorem1} can (to some extent) exploit the signals transmitted in the DIRS-jammed channel to improve the SJNR of each LU. The work in~\cite{ZFBF} showed that maximising the LU signal power in the low power domain can achieve near-optimal performance. In practice, many MU-MISO systems using low-order modulations, such as quadrature phase shift keying (QPSK), work in the low transmit power domain.

On the other hand, inter-user interference (IUI) dominates the noise for high transmit power~\cite{ZFBF}. Although the proposed anti-jamming precoder can to some extent exploit the signals transmitted in the DIRS-jammed channel, it also amplifies the interference due to the leakage from the DIRS-jammed channel. As a result, the average rate per LU resulting from the anti-jamming precoder is weaker than than that for W/O Jamming.
Compared to the rates archived by W/FPJ, the results obtained for the proposed anti-jamming algorithm are always better. For example, when the transmit power per LU is -3 dBm, the jamming impact of the DIRS-based FPJ is more serious than that of AJ with -10 dBm jamming power. However, using the proposed anti-jamming precoder, the DIRS-based jamming attacks are mitigated, and are even weaker than the jamming attacks launched by the AJ with -10 dBm jamming power.

\begin{figure}[!t]
\centering
 \begin{minipage}{0.96\linewidth}
     \centering
     \includegraphics[width=0.96\linewidth]{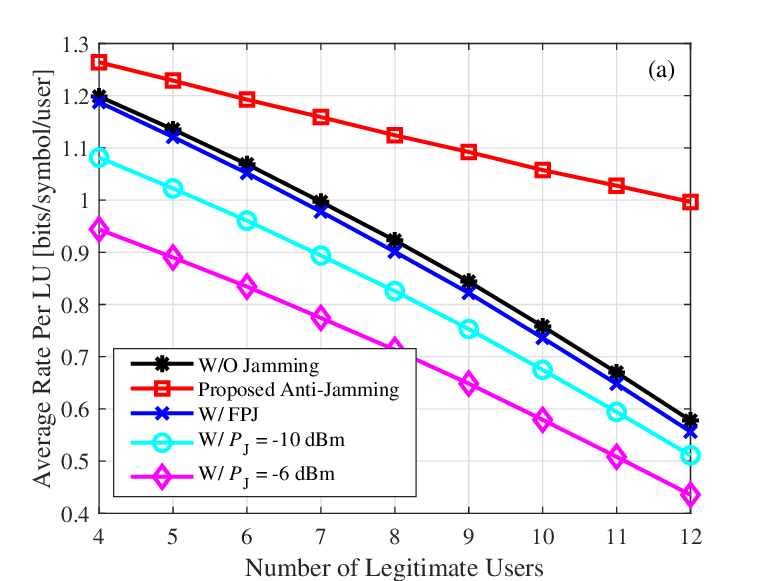}
     \label{Resfig2a}
 \end{minipage}

    \begin{minipage}{0.96\linewidth}
     \centering
     \includegraphics[width=0.96\linewidth]{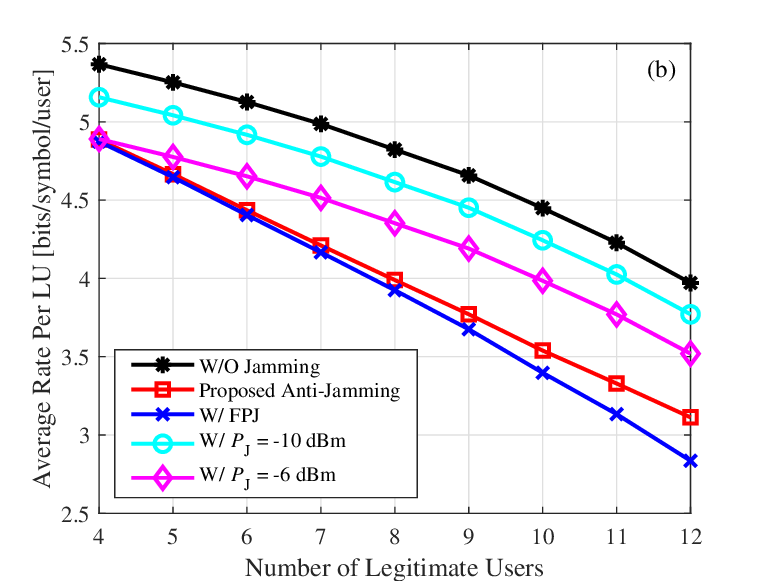}
     \label{Resfig2b}
 \end{minipage}
\caption{Number of legitimate users (LUs) vs. average rate per LU for (a) -12 dBm and (b) 3 dBm transmit power per LU.}
	\label{Resfig2}
\end{figure}

\begin{figure}[!t]
\centering
 \begin{minipage}{0.96\linewidth}
     \centering
     \includegraphics[width=0.96\linewidth]{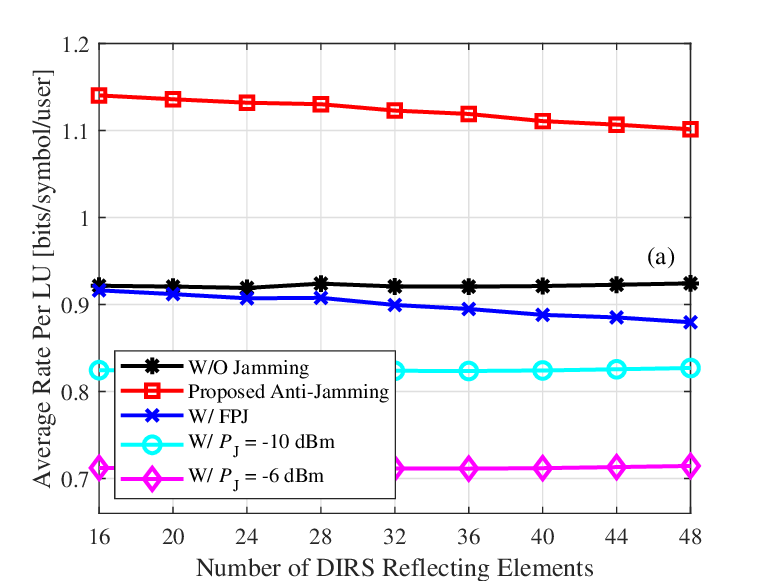}
     \label{Resfig3a}
 \end{minipage}

    \begin{minipage}{0.96\linewidth}
     \centering
     \includegraphics[width=0.96\linewidth]{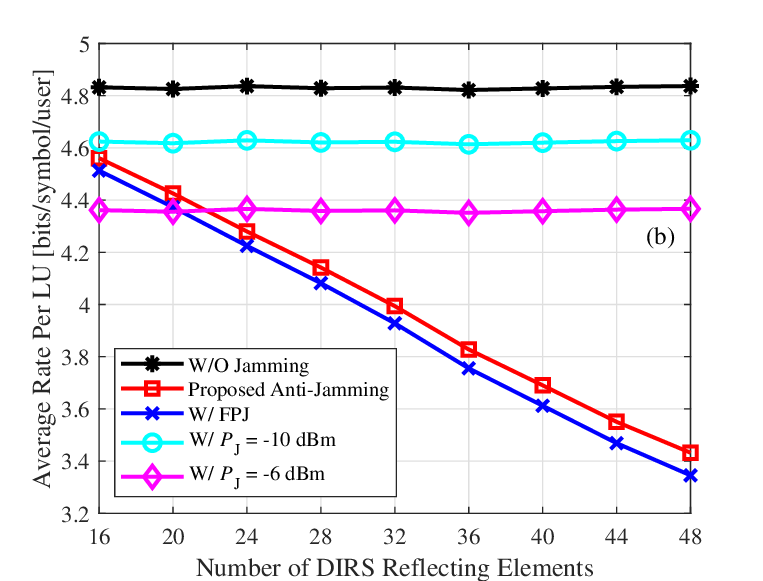}
     \label{Resfig3b}
 \end{minipage}
\caption{Number of DIRS reflecting elements vs. average rate per LU for (a) -12 dBm  and (b) 3 dBm transmit power per LU.}
	\label{Resfig3}
\end{figure}

Fig.~\ref{Resfig2} shows the average rate per LU versus the number of LUs. The results for low transmit power ($p_k = -12$ dBm) and high transmit power ($p_k = 3$ dBm) are depicted in Fig.~\ref{Resfig2} (a) and Fig.~\ref{Resfig2} (b), respectively. The rates resulting from all benchmarks decrease with the number of LUs due to the increase in IUI and the decrease in available MIMO gain.
As mentioned above, the average rate per LU resulting from the anti-jamming precoder is higher than that resulting from W/O Jamming in the low power domain. From Fig.~\ref{Resfig2} (a), one can see that the gap in average rate between the Proposed Anti-Jamming and W/FPJ increases with the number of LUs. As the number of LUs increases, the gain generated from the jammed channel becomes more significant due to the anti-jamming precoder.

In addition, the difference between the rate of W/O Jamming and the rate obtained with active jamming attacks gradually decreases as the number of LUs increases. This is due to the fact that the increase in IUI detracts from the rates, while at the same time weakening the impact of AJ. However, a unique property of the DIRS-based FPJ~\cite{DIRSVT,DIRSTWC} is that its jamming impact does not decrease as the number of LUs increases, but actually becomes more severe. Fortunately, as can be seen from Fig.~\ref{Resfig2}, the jamming mitigation generated by the anti-jamming precoder becomes more effective as the number of LUs increases.

Fig~\ref{Resfig3} displays the average rate per LU versus the number of DIRS reflecting elements for low transmit power ($p_k = -12$ dBm) and high transmit power ($p_k = 3$ dBm), respectively. Although the jamming impact of the DIRS-based FPJ becomes more severe as the number of DIRS reflecting elements increases, the proposed anti-jamming precoder always achieves higher average rates compared to the results of W/FPJ at both low and high transmit power.
\section{Conclusions}\label{Conclu}
In this paper, to address the significant threats posed by DIRS-based FPJ, a novel anti-jamming precoder has been developed that can be implemented using only the statistical characteristics of the DIRS-jammed channel instead of the instantaneous CSI.
We have showed that the elements of the DIRS-jammed channel follow a complex Gaussian distribution with zero mean and variance ${{{\mathscr{L}}\!_{{\rm G}}}{{\mathscr{L}}\!_{{\rm I},k}}{N\!_{\rm D}}\delta^2 }$. Based on this derived statistical characteristic, we have developed an anti-jamming precoder that can achieve the maximum SJNR. In particular, for an MU-MISO system operating with low power, the proposed anti-jamming precoder causes the DIRS-based FPJ to not only fail to jam the LUs, it actually improves the SJNRs of the LUs due to the additional DIRS-jammed channel it provides.
\appendices
\section{Proof of Proposition~\ref{Proposition1}}\label{AppendixA}
The element ${\left[ {{\bf H}_{\rm{D}}^{DT}} \right]_{k,n}}$ can be written as
\begin{alignat}{1}
\nonumber
{\left[ {{\bf H}_{\rm{D}}^{DT}} \right]_{k,n}}  =& \sqrt {\!\frac{{\varepsilon} {{\mathscr{L}}\!_{{\rm G}}}{\!{\mathscr{L}}\!_{{\rm I},k}} N\!_{\rm D}}{{1\!+\!\varepsilon} }} \!\!\left( \!{\frac{\sum\limits_{r = 1}^{{N_{\rm{D}}}} {\!{ {\left[ {{{\widehat { \boldsymbol{h} }}_{{\rm{I}},k}}} \right]}_r}\!{\beta\!_{D\!T,r}\!(t)}{e^{j{\varphi\!_{D\!T,r}}(t)}} }}{\sqrt{\!N_{\rm D}}}} \right.\!,\\
\nonumber
&\left. {\times {{e^{ \!- \!j\!\frac{2\pi}{\lambda}\left(\!{D_n^r}\!-{D_n}\right)}}}} \!\right) \!+\! \sqrt {\!\frac{{{\mathscr{L}}\!_{{\rm G}}}{{\mathscr{L}}\!_{{\rm I},k}} N\!_{\rm D}}{{1\!+\!\varepsilon} }} \!\left(\!{ {\frac{\sum\limits_{r = 1}^{{N_{\rm{D}}}} {{\!{\left[ {{{\widehat { \boldsymbol{h} }}_{{\rm{I}},k}}} \right]}_r}}}{\sqrt{N_{\rm D}}}} } \right.\\
&\times \left. { {\!\beta\!_{D\!T,r}(t)}{e^{j{\varphi\!_{D\!T,r}}(t)}}{\left[{\bf{G}}^{{\rm{NLOS}}}\right]_{r,n} } } \right),
\label{HDele}
\end{alignat}
where ${{\left[ {{{\widehat { \boldsymbol{h} }}_{{\rm{I}},k}}} \right]}_r}$ represents the $r$-th elements of ${{{\widehat { \boldsymbol{h} }}_{{\rm{I}},k}}}$.
Conditioned on the fact that the variables ${{\bf H}_{\rm{I}}}$, $\boldsymbol{\varphi}\!_{D\!T}(t)$, and $\bf G$ are independent, we have the following expectations:
\begin{alignat}{1}
&\mathbb{E}\left[a_r\right]\! =\! {\mathbb{E}\!\left[{ { {{ {[ {{{\widehat { \boldsymbol{h} }}_{{\rm{I}},k}}} ]}_r}{ \beta\!_{D\!T,r}(t)}{e^{j{\varphi\!_{D\!T,r}}(t)}} }} {e^{ \!- \!j\!\frac{2\pi}{\lambda}\left(\!{D_n^r}\!-{D_n}\right)}}  }\right]} \!=\! 0, \label{HDeleexpectLoS} \\
&\mathbb{E}\left[b_r\right]\! =\! {\mathbb{E}\!\left[ { {{ {[ {{{\widehat { \boldsymbol{h} }}_{{\rm{I}},k}}} ]}_r}{ \beta\!_{D\!T,r}(t)}{e^{j{\varphi\!_{D\!T,r}}(t)}} }}{[{\bf{G}}^{{\rm{NLOS}}}]_{r,n} }  \right]} \!=\! 0,
\label{HDeleexpectNLoS}
\end{alignat}
where $r=1,2,\cdots,N\!_{\rm D}$. Furthermore, the variances of $a_r$ and $b_r$ are given by
\begin{equation}
{\rm{Var}}\left[a_r\right] =  {\rm{Var}}\left[b_r\right] = \frac{{\sum\limits_{i = 1}^{{2^b}} {\kappa _i^2} }}{{{2^b}}} = \delta^2, r = 1,2,\cdots,N_{\rm D}.
\label{HDeleVars}
\end{equation}

According the central limit theorem, the random variables $\sum\nolimits_{r = 1}^{{N_{\rm{D}}}} {\frac{{{a_r}}}{{\sqrt {{N_{\rm{D}}}} }}}$ and $\sum\nolimits_{r = 1}^{{N_{\rm{D}}}} {\frac{{{b_r}}}{{\sqrt {{N_{\rm{D}}}} }}}$ converge in distribution to a
normal $\mathcal{CN}\left( {0,  \delta^2} \right)$ as $N_{\rm D} \to \infty$. Consequently, we have that
\begin{equation}
{\left[ {{\bf H}_{\rm{D}}^{DT}} \right]_{k,n}} \mathop  \to \limits^{\rm{d}}  \mathcal{CN}\!\left( {0,  {{{\mathscr{L}}\!_{{\rm G}}}{{\mathscr{L}}\!_{{\rm I},k}}{N\!_{\rm D}}\delta^2 } } \right).
\label{HDeleDis2}
\end{equation}

\end{document}